\documentstyle[12pt]{article}

\begin{document}

\newcommand{\eq}[1]{eq.~(\ref{#1})}
\newcommand{\eqs}[2]{eqs.(\ref{#1},\ref{#2})}
\newcommand{\eqss}[3]{eqs.(\ref{#1},\ref{#2},ref{#3})}
\newcommand{\eqsss}[2]{eqs.(\ref{#1}--\ref{#2})}
\newcommand{\Eq}[1]{Eq.~(\ref{#1})}
\newcommand{\Eqs}[2]{Eqs.(\ref{#1},\ref{#2})}
\newcommand{\Eqsss}[2]{Eqs.(\ref{#1}--\ref{#2})}
\newcommand{\fig}[1]{Fig.~\ref{#1}}
\newcommand{\figs}[2]{Figs.\ref{#1},\ref{#2}}
\newcommand{\figss}[3]{Figs.\ref{#1},\ref{#2},\ref{#3}}
\newcommand{\beq}{\begin{equation}}
\newcommand{\eeq}{\end{equation}}
\newcommand{\e}{\varepsilon}
\newcommand{\ee}{\epsilon}
\newcommand{\la}[1]{\label{#1}}
\def\bftau{\mbox{\boldmath$\tau$}}

\vspace{2.5cm}
\begin{center} \Large
ON SEMI-CLASSICAL PION PRODUCTION IN HEAVY ION COLLISIONS
\end{center}
\vspace{2cm}

\begin{center} \large Jean-Paul Blaizot  \\
{\small\it Service de Physique Th\'{e}orique,\\ Centre d'Etudes de Saclay,\\
F-91191 Gif-sur-Yvette, France} \\ and \\ \large Dmitri Diakonov
\footnote{Alexander von Humboldt Forschungspreistr\"{a}ger} \\ {\small\it
St.Petersburg Nuclear Physics Institute,\\ Gatchina, St.Petersburg 188350,
Russia} \end{center}
\vspace{1cm}
\thispagestyle{empty}
\vspace{2cm}
\abstract{
In high energy heavy ion collisions the pion multiplicity is large, and one
might expect that pions are radiated semi-classically. The axial symmetry
of the collision and approximately zero isotopic spin of the colliding
nuclei result then in peculiar isotopic spin -- azimuth correlations
of the produced pions. These correlations are easy to test -- and should be
tested

\newpage

    Central collisions of heavy ions at high energies are expected to
produce many pions (see for instance \cite{Satz92}), so many that the
typical number of pion quanta per unit cell of phase space may
substantially exceed unity.  Under such circumstances, one may speculate
about  the possibility that, in the early stages of the collisions, a
classical pion field develops, before subsequently decaying into
pions \cite{Anselm,Bjorken93,krz}. One interesting possibility showing how
this could happen following chiral symmetry restoration has been
investigated recently \cite{Wilczek93}.  Independently of the detailed
dynamics, the formation of an intermediate classical pion field  could
result in  certain {\it correlations} $\;$ in the production of pions with
definite isospin components.  The main point of this letter is to show
that, indeed, rather peculiar correlations can be indicated from symmetry
considerations only -- the role of dynamics is to guarantee that the pion
radiation is so strong that it can be treated semi-classically; testing the
correlations is a way to check that hypothesis. These correlations, quite
distinct from the much studied Bose Einstein correlations, find their
origin in the special correlations which develop between spatial and
isospin coordinates in the solution of the non linear field equations
obeyed by  classical pion fields.

Mathematically, one can write the amplitude of $N$ pions' production with
the help of the Lehmann--Symanzik--Zimmermann formula, where the
$N$-point pion Green function is presented through the functional integral
over the pion fields:

\[{\cal A}^{a_1 \ldots a_N}(k_1 \ldots k_N) =
\lim_{k^2_n\rightarrow m_{\pi}^2}\;
\int\! D\pi^a \int\! DJ^a\; W[J]\;
\exp{\left(iS[\pi]+i\int d^4x \pi^aJ^a\right)}\]
\beq \cdot \prod_n^N
\int \! d^4 x_n e^{ik_n x_n}
(-\partial^2_{x_n}-m_{\pi}^2) \pi^{a_n}(x_n).
\la {LSZ}\eeq

Here $S[\pi]$ is the effective pion action and $J$ is the source formed
by the colliding nuclei; one has to integrate over the sources with
some weight functional $W[J]$ which summarizes the dynamics of the
collision. If the source $J$ is in a sense ``large", the functional integral
can be evaluated in the saddle-point approximation, where the saddle-point
pion field is the solution of the equations of motion,

\beq
\frac{\delta S[\pi]}{\delta \pi^a(x)}+J^a(x)=0,
\la{eqm}\eeq
supplemented with the radiation condition at large distances, which at
$m_\pi\rightarrow 0$ reads

\beq
\frac{\partial \pi^a}{\partial t}+\frac{\partial \pi^a}{\partial r}=0.
\la{radcond}\eeq

Let us denote the solution of these eqs. as $\pi^a({\bf r},t)$. In the
leading WKB approximation one replaces the pion fields everywhere in
\eq{LSZ} by this saddle-point field. In the next approximation quantum
fluctuations about the classical field $\pi^a({\bf r},t)$ should be
also taken into account. We do not discuss dynamical questions in
this letter but concentrate solely on general symmetry considerations. We
use two assumptions:

1) the collision is axially-symmetrical;

2) the collision is isotopically-symmetrical;
that would be an exact statement in case of the zero isospin of
the colliding nuclei and is anyhow a good approximation for pions produced
in the central rapidity region since a transfer of any quantum numbers over
a long rapidity range $\Delta y$ is suppressed at least as
$(\Delta y)^{-1}$.

The first assumption means that the saddle-point pion radiation
field can be sought in the axially-symmetrical
``hedgehog" form:
\beq \pi^a({\bf r},t)=O^{ai}\left({\bf n}_{\perp
i}P_{\perp}(\rho,z,t)+{\bf n}_{\parallel i}P_{\parallel}(\rho,z,t)\right)
\la{axsym}\eeq
where $P_{\perp , \parallel}$ are functions of the distance $\rho$ from the
beam axis, of the distance $z$ from the collison point and of time $t$;
${\bf n}_{\parallel (\perp)}$ are unit vectors parallel (perpendicular)
to the beam axis and $O^{ai}$ is an arbitrary $3\times 3$ orthogonal
matrix.  (In principle, axial symmetry does not contradict higher harmonics
in the transverse plane with ${\bf n}_{\perp}$ replaced by a unit vector
with the components $(\cos m\phi, \; \sin m\phi)$ where $\phi$ is the
azimuthal angle and $m$ is an integer.  We shall work with $m=1$, i.e.
assume that ${\bf n}_{\perp}$ points in the radial direction, and introduce
the arbitrariness in the choice of $m$ only in our final results). The type of
correlations between space and isospin degrees of freedom is a familiar feature
of solutions of non linear field equations satisfied by the pion fields, as
illustrated for example in the Skyrme model (see, e.g. \cite{Bala91}). The
second assumption means that the saddle point is degenerate in the global
isospin rotation $O^{ai}$, so that the functional integral in \eq{LSZ}
reduces to the integration over all possible orientations of the pion field
in the isospin space. The action and the source-weight functionals are
taken at the saddle-point values of $J^a$ and $\pi^a$ and provide an
overall normalization factor $\surd \cal N$ which may be a function of the
total 4-momentum of the pions $P_\mu$, but now we are not interested in
this factor.

At large distances / time  the isospin source $J^a$ dies out and one can
also neglect the non-linearity of the pion effective action. Therefore,
\eq{eqm} reduces to the free Klein--Gordon eq. at large distances, and
we are guaranteed that the Fourier transform of \eq{axsym}
has a pole at $k_0^2-k_z^2-{\bf k}_{\perp}^2=m_{\pi}^2$, which cancels
out in the LSZ leg amputation procedure (see \eq{LSZ}). We have
therefore:

\[ \lim_{k^2\rightarrow m_{\pi}^2} \int \! d^4 x e^{ikx}
(-\partial^2_{x}-m_{\pi}^2) \pi^{a}(x)=
O^{ai}\left({\bf k}_{\perp i}F_{\perp}(k_\perp,k_\parallel)
+{\bf k}_{\parallel i}F_{\parallel}(k_\perp,k_\parallel)\right)\]
\beq\equiv O^{ai}F_i({\bf k})
\la{Fpi}\eeq
where $F_{\perp,\parallel}(k_\perp,k_\parallel)$
are related through Fourier transformation to $P_{\perp,\parallel}$ of
\eq{axsym}. Since $P_{\perp,\parallel}$ are real the functions $F_i({\bf
k})$ are purely imaginary, with $F_i^{\star}({\bf k}) = F_i({-\bf k})
= -F_i({\bf k})$.

Squaring the amplitudes, summing over the isospin $a=1,2,3$ of the pions
and multiplying by the phase space factor, one gets for the $N$ pion
production cross-section:
 \[\sigma^{(N)}=\frac{{\cal N}}{N!}\int \! dO_1 dO_2
\int \!\prod_{n=1}^N \frac{d^4k_n}{(2\pi)^4} 2\pi \delta_+(k^2_n-m_{\pi}^2)
(2\pi)^4\]
\beq \cdot O_1^{ai}F_i({\bf k}_n)O_2^{aj}F_j^\star({\bf k}_n)
\delta^{(4)}(P_\mu-\Sigma k_{n\mu}),
\la{sigmaN}\eeq
where $k_{n\mu}$ are individual 4-momenta of the produced pions, $O_{1,2}$
are isospin orientation of the pions in the amplitude and the conjugate
amplitude, respectively, $P$ is the total 4-momentum of the produced
pions; the factorial accounts for the identical particles and
$\delta_+(k^2_n-m_{\pi}^2)=\delta(k^2_n-m_{\pi}^2)\theta(k_0)$.

Writing the 4-momentum conservation restriction as
\beq
(2\pi)^{4}\delta^{(4)}(P-\sum k_n)
= \int\! d^4R \!\,\, e^{i(P\cdot R)-i\sum(k_n\cdot R)},
\la{momentumcons}\eeq
we get factorized integrals over the momenta of produced pions:
\beq
\int \! \frac{d^4k}{(2\pi)^4} 2\pi \delta_+(k^2-m_\pi^2) e^{-i(k\cdot
R)}F_i({\bf k}) F_j^\star({\bf k}) = {\cal F}_{ij}(R_0, {\bf R}).
\la{Four}\eeq

This tensor is further on contracted with the {\it relative} orientation
matrix $O_{12}^{ij} \equiv O_1^{ai}O_2^{aj}=(O_1^TO_2)^{ij}$. We note that
in integrating over the $SO(3)$ rotations one can use the Haar measure
property, $dO = d(CO) = d(OC)$, where $C$ is an arbitrary orthogonal
matrix, so that
\beq
\int\!\int dO_1 dO_2 = \int\!\int dO_1 dO_{12}\; ; \;\;\;\;\; \int dO = 1.
\la{Haar}\eeq

The total cross section being a sum of $\sigma^{(N)}$ over $N$ becomes
thus a series for an exponent, and we get
\beq
\sigma^{tot} = {\cal N}\int dO_{12}
\int d^4 R \exp \left[ i(P\cdot R) +
{\cal F}_{ij} (R_0, {\bf R}) O_{12}^{ij} \right].
\la{sigmatot}\eeq

Let us use the frame in which the total momentum of the produced pions
is zero, ${\bf P}=0, \; P_0=E$ where $E$ is the total energy of the pions.
Since both $E$ and the function $\cal F$, proportional to the probability
of the pion production, are presumably large, one can integrate over
$R_0$, ${\bf R}$ and $O_{12}$ by the saddle-point method.
Let us parametrize the relative orientation matrix $O_{12}$ in terms of a
unit 4-vector $u_\mu \; \left (u_\mu^2 = 1, \tau_\mu = (1, -i {\bftau})
\right )$,
\beq
O_{12}^{ij} = \frac{1}{2} Tr (u_\mu \tau_\mu \tau^i u_\nu
\tau_\nu^\dagger \tau^j) = (1-2{\bf u}^2)\delta{ij}+2u_iu_j+2u_0u_k\e_{ijk}.
\la{O12}\eeq

We expect the saddle point to be at
\beq
{\bf u} \approx 0,\;\;\; {\bf R}\approx 0, \;\;\; R_0 \sim \frac{1}{E}.
\la{SP}\eeq

Expanding the exponent in \eq{sigmatot} around this point, and using the
explicit form of ${\cal F}$ given by \eq{Four}, it can be shown that, at
small $R_0$, the saddle-point condition is indeed satisfied. Without
knowing the explicit form of the functions $F_i({\bf k})$ we cannot prove
that there are no other saddle-points, but we shall disregard such
possibilities. Note that the maximum at ${\bf u}=0$ corresponds to
$O_{12}^{ij}=\delta^{ij}$, i.e. to the case when the pion isospin
orientation in the conjugate amplitude is the same as in the direct
amplitude -- not an unnatural result.

We thus write the total cross section as
\beq
\sigma^{tot} \approx {\cal N} \int dR_0 \exp
\left[ i(ER_0) + {\cal F}_{ii}(R_0,{\bf 0}) \right].
\la{Total}\eeq
In what follows we shall use this relation to remove the unknown
normalization factor $\cal N$.

We next turn to the 1,2,... particle {\it inclusive} cross sections. All of
them are directly derived from \eq{sigmaN} where one skips integration over
1,2,... momenta and summation over the 1,2,... isotopic subscripts. Thus,
the 1-particle inclusive cross section is given by (we use the abbreviation
$(d{\bf k})=d^3{\bf k}/2k_0(2\pi)^3)$:
\[
\frac{d\sigma^{(A)}}{(d{\bf k})}=u_b^{(A)}u_{b^\prime}^{\star(A)}
\int dO_1 dO_2 O_1^{bi}O_2^{b^\prime j} F_i({\bf k}) F_j^\star({\bf k})\]
\beq \cdot
{\cal N} \sum_{N=1}^\infty \frac{1}{(N-1)!} \int \prod_n^{N-1}(d{\bf k}_n)
\left(F({\bf k}_n)O_{12}F^\star({\bf k}_n)\right) (2\pi)^4
\delta^{(4)}(P-k-\sum k_n).
\la{1incl1}\eeq
Here $u_b^{(A)}$ are the isospin "polarization" vectors,

\beq
u^{(\pi^0)}_b=\left( \begin{array}{c}
0 \\ 0 \\ 1 \end{array} \right)_b,\;\;\; u_b^{(\pi^+)}=\frac{1}{\surd
2}\left( \begin{array}{c} 1 \\ i \\ 0 \end{array} \right)_b, \;\;\;
u_b^{(\pi^-)}=\frac{1}{\surd 2}\left( \begin{array}{c} 1 \\ -i \\ 0
\end{array} \right)_b,
\la{polar}\eeq
to be contracted with the isospin orientation matrices $O_{1,2}$.
The superscript $A$ refers to the isospin component of the observed pion;
there is no summation on $A$.

The 2-particle inclusive cross section for production of the pion of sort
$A_1$ with momentum ${\bf k}_1$ and of the pion of sort $A_2$ with momentum
${\bf k}_2$ is

\[
\frac{d\sigma^{(A_1A_2)}}{(d{\bf k}_1)(d{\bf k}_2)} =
u_b^{(A_1)}u_{b^\prime}^{\star(A_1)}
u_c^{(A_2)}u_{c^\prime}^{\star(A_2)}
\]
\[
\cdot \int dO_1 dO_2 O_1^{bi}O_2^{b^\prime j}
F_i({\bf k}_1) F_j^\star({\bf k}_1) O_1^{ck}O_2^{c^\prime l}
F_k({\bf k}_2) F_l^\star({\bf k}_2)
{\cal N} \sum_{N=2}^\infty \frac{1}{(N-2)!}
\]
\beq
\cdot  \int\prod_n^{N-2}
(d{\bf k}_n) \left (F({\bf k}_n)O_{12}F^\star({\bf k}_n)\right ) (2\pi)^4
\delta^{(4)}(P-k_1-k_2-\sum k_n),
\la{2incl1}\eeq
and so on. Writing the 4-momentum conservation $\delta$ function with the
help of an auxiliary integral as in \eq{momentumcons}, we again obtain
the exponential series, so that
\[
\frac{d\sigma^{(A)}}{(d{\bf k})}=u_b^{(A)}u_{b^\prime}^{\star(A)}
\int dO_1 dO_{12} O_1^{bi}O_1^{b^\prime k} O_{12}^{kj}F_i({\bf k})
F_j^\star({\bf k})
\]
\beq
\cdot {\cal N} \int d^4 R \exp \left[ i(P\cdot R) - i(k \cdot R) +
{\cal F}_{ij} (R) O_{12}^{ij} \right]
\la{1incl2}\eeq
where we used $O_2^{b^\prime j}=O_1^{b^\prime k} O_{12}^{kj}$. If the
momentum of the observed pion is negligible as compared to the total
momentum $P$ of the pions and if the integration  over $O_{12}$
and $R$ is performed about the presumably steep saddle point given by
\eq{SP}, we obtain:

\[
\frac{d\sigma^{(A)}}{(d{\bf k})} \simeq
 u_b^{(A)}u_{b^\prime}^{\star(A)} \int dO_1 O_1^{bi} O_1^{b^\prime j}
F_i({\bf k})F_j^\star({\bf k}) \sigma^{tot}
\]
\beq =
u_b^{(A)}u_{b^\prime}^{\star(A)} \frac{1}{3} \delta^{b b^\prime}
\delta^{ij} F_i({\bf k})F_j^\star({\bf k}) \sigma^{tot} =
\frac{1}{3} F_i({\bf k})F_i^\star({\bf k}) \sigma^{tot}.
\la{1incl3}\eeq

We see that the inclusive cross section is directly related to the
square of the Fourier transform of the classical pion radiation field, --
a most natural result. Also naturally, we find identical cross sections for
$\pi^+, \; \pi^-\;$ and $\pi^0$ production. The average multiplicity is
obtained by integrating the inclusive cross section over ${\bf k}$ and
summing over $A=\pi^+,\pi^-,\pi^0$:
\beq
\langle N \rangle \approx \int (d{\bf k}) F_i({\bf k})F_i^\star({\bf k})
\approx {\cal F}_{ii}(R_\mu=0).
\la{multipl}\eeq
It should be kept in mind that if this integral is not convergent by itself
at large ${\bf k}$ it should be cut at least at $P$ in accordance with a
more precise \eq{1incl2}.

In \eq{1incl2} we used the following formula for averaging over isospin
rotations:
\beq
\int dO O^{bi} O^{b^\prime k} = \frac{1}{3} \delta^{bi} \delta^{b^\prime k}.
\la{av1}\eeq
To calculate the 2-particle inclusive cross section we need a formula for
averaging over four matrices:
\[
\int dO O^{bi} O^{b^\prime j} O^{ck} O^{c^\prime l} =
\frac{1}{30}\delta^{ij}\delta^{kl}(4\delta^{bb^\prime}\delta^{cc^\prime}
-\delta^{bc}\delta^{b^\prime c^\prime}
-\delta^{bc^\prime}\delta^{b^\prime c})\]
\[
+ \frac{1}{30}\delta^{ik}\delta^{jl}(-\delta^{bb^\prime}\delta^{cc^\prime}
+4\delta^{bc}\delta^{b^\prime c^\prime}
-\delta^{bc^\prime}\delta^{b^\prime c})\]
\beq
+ \frac{1}{30}\delta^{il}\delta^{jk}(-\delta^{bb^\prime}\delta^{cc^\prime}
-\delta^{bc}\delta^{b^\prime c^\prime}
+4\delta^{bc^\prime}\delta^{b^\prime c}).
\la{av2}\eeq
One can check this formula by applying various contractions and reducing it
to \eq{av1}; an alternative method is to note that $O^{bi}$ is a Wigner $D$
function for isospin 1, and using the Clebsch--Gordan machinery.

Starting from \eq{2incl1} and repeating the same steps as above we get
for the 2-particle inclusive cross section:

\[
\frac{d\sigma^{A_1A_2}}{(d{\bf k}_1)(d{\bf k}_2)} =
u_b^{(A_1)}u_{b^\prime}^{\star(A_1)}u_c^{(A_2)}u_{c^\prime}^{\star(A_2)}
\]
\[
\cdot \frac{1}{30}\left[2\delta^{bb^\prime}\delta^{cc^\prime}(2V-W)
+(\delta^{bc}\delta^{b^\prime c^\prime}
+\delta^{bc^\prime}\delta^{b^\prime c})(-V+3W)\right]\sigma^{tot}
\]
\beq
=\frac{\sigma^{tot}}{30}
\left\{ \begin{array}{lll} 3V+W
\\ 4V-2W \\ 2V+4W \end{array} \right. =
\left\{ \begin{array}{lll} \sigma^{(1)} \\ \sigma^{(2)}\\ \sigma^{(3)}
\end{array} \right.
\la{2incl2}\eeq
where we have introduced the abbreviation:
\beq V=|F_i({\bf k}_1)|^2 |F_j({\bf k}_2)|^2, \qquad
W=|F_i({\bf k}_1) F_i({\bf k}_2)|^2.
\la{VW}\eeq
The upper line in \eq{2incl2} (case 1) refers to the $\pi^+\pi^-,\pi^-\pi^+,
\pi^+\pi^+$ or $\pi^-\pi^-$ production, the second line (case 2) refers to
four other combinations, $\pi^+\pi^0,\pi^-\pi^0,$ $\pi^0\pi^+,\pi^0\pi^-$,
while the last line (case 3) corresponds to the $\pi^0\pi^0$ production.
It should be noted that, if the two observed pions are not identical,
they are distinguished by the momenta ${\bf k}_1,{\bf k}_2$. The three
possibilities in \eq{2incl2} reflect three possible isospin states
$(T = 0, 1, 2)$ which can be formed by a pair of pions. Since, however,
the three cross sections are expressed through only two functions, we get a
relation:
\beq
\sigma^{(2)} + \sigma^{(3)} = 2 \sigma^{(1)}.
\la{relation}\eeq

Let us now investigate \eq{2incl2}. If one sums up all 9 possible
combinations of pion pairs, one gets
\beq
\frac{d\sigma^{all}}{(d{\bf k}_1)(d{\bf k}_2)} =
|F_i({\bf k}_1)|^2 |F_j({\bf k}_2)|^2 \sigma^{tot}
\la{all}\eeq
which is independent of the angle between the two pions. Further on,
integrating \eq{all} over the momenta ${\bf k}_{1,2}$ and recalling
\eq{multipl} for the average multiplicity, one finds
\beq
\langle N(N-1) \rangle =\frac{1}{\sigma^{tot}}
\int (d{\bf k}_1)(d{\bf k}_2)\frac{d\sigma^{all}}{(d{\bf k}_1)(d{\bf k}_2)}
=\langle N \rangle ^2,
\la{dispersion}\eeq
which is the dispersion law of the Poisson distribution. It should be
stressed though that for charged or neutral pions separately there is no
Poisson distrubution! Also, we would expect a deviation from the Poisson
distribution at the end point of the spectrum where the 4-momentum
conservation law from a more accurate \eq{2incl1} imposes additional
correlations.

The structure denoted as $W$ depends on the azimuthal angle between the two
pions,
\beq
W=\left[\cos(\phi_1-\phi_2)k_1^\perp k_2^\perp F_\perp (k_1)F_\perp (k_2)+
k_1^\parallel k_2^\parallel F_\parallel (k_1)F_\parallel (k_2)\right]^2,
\la{W}\eeq
while
\beq
V=\left[(k_1^{\perp})^2 F_\perp^2 (k_1))+ (k_1^{\parallel})^2F_\parallel^2
(k_1)
\right] \left[(k_2^{\perp})^2 F_\perp^2 (k_2))+
(k_2^{\parallel})^2F_\parallel^2
(k_2)\right]. \la{V}\eeq
Out of the three cross sections mentioned in \eq{2incl2} one can construct
angle-dependent and -independent combinations:
\[
2\sigma^{(3)}-\sigma^{(2)} =
4\sigma^{(1)}-3\sigma^{(2)} = \frac{\sigma^{tot}}{3}W({\bf k}_1,{\bf k}_2),
\]
\beq
2\sigma^{(1)}+\sigma^{(2)} =
2\sigma^{(2)}+\sigma^{(3)} = \frac{\sigma^{tot}}{3}V({\bf k}_1,{\bf k}_2).
\la{comb}\eeq

\Eqsss{2incl2}{comb} summarize our result for double-inclusive pion
production.  However, it might be useful to make a prediction which is
independent of the dynamics hidden in the Fourier-transformed pion fields
$F_{\perp,\parallel}$. To this end we restrict ourselves to pions with zero
rapidity, i.e. to the case $k_1^\parallel=k_2^\parallel=0$, so that $W/V =
\cos^2(\phi_1-\phi_2)$. This quantity is obtained by taking the ratio of
differential double-inclusive cross sections, say,
\beq
\frac{4\sigma^{(1)}-3\sigma^{(2)}}{2\sigma^{(2)}+\sigma^{(3)}} =
\cos^2(\phi_1-\phi_2).
\la{ratio1}\eeq
Another way to isolate the azimuthal angle dependence is to normalize to
the single-inclusive cross sections. For example, we predict at
$k_{1,2}^\parallel=0$:
\beq
\left(\sigma^{tot} \frac{d\sigma^{\pi^+\pi^-}}{(d{\bf k}_1)(d{\bf k}_2)}
-\frac{9}{10} \frac{d\sigma^{\pi^+}}{(d{\bf k}_1)}
\frac{d\sigma^{\pi^-}}{(d{\bf k}_2)} \right) /
\left(
\frac{d\sigma^{\pi^+}}{(d{\bf k}_1)}
\frac{d\sigma^{\pi^-}}{(d{\bf k}_2)} \right)
= \frac{3}{10} \cos^2(\phi_1-\phi_2).
\la{ratio2}\eeq

At this point an experimentalist may derive correlations for his own
favourite (charged or neutral) pairs of pions. Let us recall finally that the
axial symmetry does not contradict higher harmonics in the transverse
plane, with the replacement $\cos(\phi_1-\phi_2)
\rightarrow \cos m(\phi_1-\phi_2)$ where $m$ is an integer.
\vspace{.5cm}

D.D. would like to thank Maxim Polyakov and Michal Praszalowicz for a
helpful discussion, the Institute for Theoretical Physics-II of the Ruhr
University at Bochum for hospitality during the completion of this paper
and the A.v.Humboldt - Stiftung for a support.


\vskip 1cm



\begin{thebibliography}{}


\bibitem{Satz92}
H.Satz, {\it Nucl. Phys.} {\bf A544} (1992) 371c

\bibitem{Anselm}
A.A.~Anselm, {\it Phys. Lett.} {\bf 217B} (1989) 169; A.A.~Anselm and
M.~Ryskin, {\it Phys. Lett.} {\bf 266B} (1991) 482

\bibitem{Bjorken93}
J.D.~Bjorken, {\it Acta Physica Polonica} {\bf B23} (1992) 561

\bibitem{krz}
J.P.~Blaizot and A.~Krzywicki, {\it Phys. Rev.} {\bf D46} (1992) 246

\bibitem{Wilczek93}
K.~Rajagopal and F.~Wilczek, {\it ``Emergence of Coherent Long Wavelength
Oscillations After a Quench: Application to QCD''} PUPT-1389,
IASSNS-HEP-93/16 , March 93

\bibitem{Bala91}
A.P.~Balachandran, G.~Marmo, B.S.~Skagerstam and A.~Stern,
{\it ``Classical Topology and Quantum States''}, World Scientific (1991)

\end{thebibliography}
\end{document}